\documentclass{article}
\usepackage{amsmath}
\usepackage{amssymb}
\usepackage{graphicx}
\begin{document}

\begin{center}
\LARGE
\textbf{Quantum teleportation and Grover's algorithm without the wavefunction}
\\[0,5 cm]
\normalsize
Gerd Niestegge
\\[0,5 cm]
\footnotesize
Zillertalstrasse 39, 81373 M\"unchen, Germany

gerd.niestegge@web.de
\\[0,5 cm]
\end{center}
\normalsize

\begin{abstract}
In the same way as the quantum no-cloning theorem and 
quantum key distribution in two preceding papers,
en\-tanglement-assisted quantum teleportation 
and Grover's search algorithm are generalized 
by transferring them to 
an abstract setting, including usual 
quantum mechanics as a special case.
This again shows that a much more general and abstract access
to these quantum mechanical features is possible
than commonly thought. 
A non-classical extension of conditional probability 
and, particularly, a very special type of 
state-independent conditional probability 
are used instead of  Hilbert spaces and wavefunctions.
\\[0,5 cm]
\\[0,5 cm]
\\[0,5 cm]
\end{abstract}
\noindent
\large
\textbf{1. Introduction}
\normalsize
\\[0,5 cm]
In the past thirty years, quantum information 
has been a wide field of extensive theoretical and experimental research.
Important topics are
the quantum no-cloning theorem \cite{dieks1982communication, wootters1982single},
quantum cryptography including particularly quantum key distribution \cite{BB84, E91},
entanglement-assisted quantum teleportation \cite{bennett1993teleporting}, 
and quantum computing with its specific quantum algorithms
like the Deutsch-Jozsa algorithm \cite{Cleve-Ekert-et-al1998, Deutsch1985, Deutsch-Jozsa1992},
Grover's search algorithm \cite{grover1996, grover1997} 
and Shor's factoring algorithm \cite{shor1994}. 

In two recent papers \cite{Nie2015PhysScrNoCloning, nie2017QKD}, 
it has been demonstrated that
the quantum no-cloning theorem 
and quantum key distribution
allow a much more general and abstract access than
commonly thought. This approach uses 
a non-classical extension of conditional probability
\cite{niestegge2001non, niestegge2008approach}
and, particularly, a very special type of 
state-independent conditional probability 
instead of Hilbert spaces and wavefunctions.

In the present paper, it is shown
that the same approach is applicable 
to two further topics of quantum information theory. 
Entanglement-assisted quantum teleportation and  
Grover's search algorithm are considered.
It is shown that these topics allow the same 
general and abstract access as the no-cloning theorem
and quantum key distribution.
This time, however, it becomes necessary 
to go a little deeper into the theory of 
the non-classical conditional probabilities.
Sequential conditionalization and the representation 
of probability conditionalization by transformations
on a certain order-unit space \cite{nie2008order-unitspace} must be considered.

A vast amount of papers on quantum teleportation and 
Grover's algorithm,
concerning theoretical studies as well as experimental set-ups,
is available by now; the pioneering
paper on quantum teleportation \cite{bennett1993teleporting} has more than
ten thousand quotations, and each one of Grover's two papers \cite{grover1996, grover1997}
more than three thousand. The present paper focuses on the theoretical foundations
and on the new general abstract access
to these quantum mechanical features.
Other general and abstract studies of teleportation and 
the search algorithm \cite{barnum2012teleportation, Lee-Selby2016}
identify mathematical conditions or 
physical principles making these features work in a 
Generalized Probabilistic Theory.
The approach presented here differs from them  
in the considered mathematical conditions or 
physical principles and, particularly, in the 
central role played by the special type of 
state-independent conditional probability.

The paper is organized as follows.
Section 2 briefly restates some material from 
Refs. \cite{brabec1979compatibility, niestegge2001non, niestegge2008approach, nie2008order-unitspace}
as far as needed in the present paper; the particular topics are:
compatibility in quantum logics, the non-classical extension of 
conditional probability and the representation of probability 
conditionalization by
transformations on a certain order-unit space.
Section 3 presents two specific assumptions
which will play an important role in the rest of the paper; 
the second one turns out to represent an interesting
property of quantum mechanics which has not been known so far.
Sections 4 and 5 contain the main results concerning the 
the new general and abstract access to
quantum teleportation and Grover's algorithm. A lemma with a longish mathematical proof, 
concerning the success probability of Grover's algorithm, is shifted to the annex.
The link to the well-known Hilbert space versions of these quantum mechanical
features is elucidated is section 6.
\\[0,5 cm]
\large
\textbf{2. Non-classical conditional probability}
\\[0,5 cm]
\normalsize
\textbf{2.1 The quantum logic}
\\[0,5 cm]
In quantum mechanics, the measurable quantities of a physical 
system are re\-presented by observables. Most simple are those 
observables where only the two values `true' and `false'
(or `1' and `0')
are possible as measurement outcome.
They are elements of a mathematical structure called \textit{quantum logic},
are usually called \textit{propositions}, 
and they are called \textit{events} in probabilistic approaches.
The elements of the quantum logic can also be understood as potential properties 
of the system under consideration. 

In this paper, the quantum logic shall be an 
orthomodular partially ordered set $E$ 
with the partial ordering $\leq$,
the orthocomplementation $'$, 
the smallest element $0$ and the largest element $\mathbb{I}$
\cite{beltrametti1984logic, beran1985orthomodular, kalmbachorthomodular, ptak1991orthomodular}.
Two elements $e,f \in E$ are called \textit{orthogonal} if $e \leq f'$ or, equivalently, $f \leq e'$.
An element $e \neq 0$ in $E$ is called an \textit{atom} if there is no element $f$ in $E$ 
with $f \leq e$ and $0 \neq f \neq e$. 
The interpretation of this mathematical terminology is as follows: 
two orthogonal elements represent mutually exclusive events, 
propositions or system properties, and $e'$ represents the negation of $e$.
\\[0,5 cm]
\textbf{2.2 Compatibility}
\\[0,5 cm]
Classical probability theory uses Boolean algebras as mathematical structure 
for the random events, and it can be expected that those subsets of $E$,
which are Boolean algebras, behave classically. Therefore, a subset $E_0$ of $E$
is called \textit{compatible} if there is a Boolean algebra $B$ with $E_0 \subseteq B \subseteq E$.
Two elements $e$ and $f$ in $E$ are called compatible, if $\left\{e,f\right\}$ forms a compatible subset.
Note that the supremum $e \vee f$ and the infimum $e \wedge f$ exist 
for any compatible pair $e$ and $f$ in $E$
and that the distributivity law 
$e \wedge (f \vee g) = (e \wedge f) \vee (e \wedge g)$ 
holds for $e,f,g$ in any compatible subset of $E$.
Any subset with pairwise orthogonal elements is compatible \cite{brabec1979compatibility}.

Two subsets $E_1$ and $E_2$ of $E$ are called \textit{compatible with each other}
if the union of any compatible subset of $E_1$ with any compatible subset of $E_2$
is a compatible subset of $E$. Note that this does not imply that $E_1$ or $E_2$ themselves 
are compatible subsets.

A subset of an orthomodular lattice 
(i.e., the supremum $e \vee f$ and the infimum $e \wedge f$ exist 
not only for the compatible, but for all pairs $e$ and $f$ in $E$) 
is compatible if each pair of elements in this subset 
is compatible. However, the pairwise compatibility of the elements
of a subset of an orthomodular partially ordered set 
does not any more imply the compatibility of this subset 
\cite{brabec1979compatibility}.

For compatible pairs, the supremum $\vee$ and the infimum $\wedge$
represent the logical or- and and-operations. For incompatible pairs, 
the supremum $\vee$ and the infimum $\wedge$ may exist 
as mathematical objects, but do not have any interpretation.
\\[0,5 cm]
\textbf{2.3 Conditional probability}
\\[0,5 cm]
The states on the orthomodular partially ordered set $E$
are the analogue of the 
probability measures in classical probability theory, and 
conditional probabilities can be defined similar to 
their classical prototype. 

A \textit{state} $\rho$ allocates the probability $ \rho(f)$ 
with $0 \leq \rho(f) \leq 1$ to each element $f \in E$, 
is additive for orthogonal elements, and $\rho(\mathbb{I})=1$.
It then follows that $\rho(f) \leq \rho(e)$ for any two elements
$e,f \in E$ with $f \leq e$.

The \textit{conditional 
probability} of an element $f \in E$ under another element $e \in E$ is the 
updated probability for $f$ after the outcome of
a first measurement has been $e$; it is denoted 
by $ \rho(f | e) $. Mathematically, it is defined by the
conditions that the map $E \ni f \rightarrow \rho(f | e)$
is a state on $E$ and that it coincides with the classical
conditional probability for those $f$ which are compatible
with $e$; this means 
$\rho(f | e) = \rho(e \wedge f) / \rho(e)$,
if $f$ is compatible with $e$.
It must be assumed that $\rho(e) \neq 0$.
\newpage
However, among the orthomodular partially ordered sets, 
there are many where no states or no conditional 
probabilities exist, or where the conditional probabilities 
are ambiguous. It shall now be assumed 
for the remaining part of this paper
that there is a state $\rho$ on $E$ with $\rho(e)\neq 0$ for each $e \in E$ with $e \neq 0$,
that $E$ possesses unique conditional probabilities, 
and that the state space of $E$ is strong (i.e., if 
$\left\{ \rho \ |\  \rho \mbox{ is a state with } \rho(f) = 1 \right\} 
\subseteq \left\{ \rho \ |\  \rho \mbox{ is a state with } \rho(e) = 1 \right\}$
holds for $e, f \in E$, then $f \leq e$). 
Note that, if $\rho$ is a state with $\rho(e) = 1$ for some element $e \in E$, 
$ \rho(f | e) = \rho(f)$ for all $f \in E$. 

For some pairs $e$ and $f$ in $E$, the conditional probability
does not depend on the underlying state; this means 
$\rho_1 (f|e) = \rho_2 (f|e)$ for all states $\rho_1$ and $\rho_2$
with $\rho_1 (e) \neq 0 \neq \rho_2 (e)$. This special conditional
probability is then denoted by $\mathbb{P} (f|e)$. 
It results solely from the algebraic structure of the 
quantum logic and, therefore, it is invariant under morphisms \cite{Nie2015PhysScrNoCloning}. 

For $e,f \in E$, 
$\mathbb{P} (f|e)$ exists and $\mathbb{P} (f|e) = p$ if and only if
$\rho(e) = 1$ implies $\rho(f) = p$ for the states $\rho$ on $E$. 
Moreover, $f \leq e$ holds for two elements $e$ and $f$ in $E$
if and only if $ \mathbb{P} (e|f) = 1$, and  $e$ and $f$ are orthogonal
if and only if $ \mathbb{P} (e|f) = 0$.
 
$\mathbb{P} (f|e)$ exists for all $f \in E$ if and only if $e$ is an \textit{atom}
(minimal element in $E$), which results in the atomic state
$\mathbb{P}_e$ defined by $\mathbb{P}_e (f) := \mathbb{P} (f|e)$.
This is the unique state allocating the probability value $1$ to the atom $e$.

The type of conditional probability, considered here, was introduced in 
Refs. \cite{niestegge2001non, niestegge2008approach}, 
to which it is referred for more information.
\\[0,5 cm]
\textbf{2.4 The order-unit space}
\\[0,5 cm]
The quantum logic $E$ generates an order-unit space $A$
(partially ordered real linear space with a specific norm; see \cite{AS01})
and can be embedded in its unit interval $\left[0,\mathbb{I}\right]$
$:=$ $\left\{a \in A : 0 \leq a \leq \mathbb{I} \right\}$ 
$=$ $\left\{a \in A : 0 \leq a\ \text{and} \left\| a \right\| \leq 1 \right\}$;
$\mathbb{I}$ becomes the order-unit, and
$e' = \mathbb{I} - e$ for $e \in E$. 
Each state $\mu$ on $E$ has a unique positive linear extension on $A$ 
which is again denoted by $\mu$.

As shown in \cite{nie2008order-unitspace, nie2012AMP}, 
for each element $e$ in $E$, there is a positive linear operator 
$U_e : A \rightarrow A$ with 
$\mu(f|e)\ \mu(e) = \mu(U_e f)$ for all $f \in E$ and all states $ \mu $.
This means that probability conditionalization is represented 
by the transformations $U_e$ on the order-unit space $A$.
If $\mu(e) = 1$, then $\mu(U_e f) = \mu(f)$ for all $f \in E$ (or briefly $\mu U_e = \mu$)
and, if $\mu(e) = 0$, then $\mu(U_e f) = 0$ for all $f \in E$ (or briefly $\mu U_e = 0$).

These transformations satisfy $U_e^{2} = U_e$.
Moreover, with $e,f \in E$, $\mathbb{P}(f|e) = p$ if and only if $U_e f = p e$. 
Furtheremore, $U_e f = U_f e = e\wedge f$ and $U_e U_f = U_f U_e = U_{e\wedge f}$ 
for any compatible pair $e$ and $f$ in $E$; this follows from Lemma 3 in \cite{Nie2010CTP}.
Particularly, $U_e e = e = U_e \mathbb{I}$, $U_e e' = 0$, and $U_e f = 0$ 
if and only if $e$ and $f$ are orthogonal.
\newpage
\noindent
\large
\textbf{3. Two assumptions}
\\[0,5 cm]
\normalsize
\textbf{3.1 The $S$-transformations}
\\[0,5 cm]
For each $e \in E$, a further  
linear operator $S_e$ on $A$ can be defined by 
$$S_e x := 2U_e x + 2 U_{e'}x - x, x \in A.$$
The above properties of $U_e$ imply that 
$S_e^{2}x = x$ for all $x$ in $A$.
This means that $S_e$ is its own inverse: $S_e = S_e^{-1}$.
Further important properties of $S_e$ are: $S_e = S_{e'}$, $S_e U_e = U_e$, $S_e U_{e'} = U_{e'}$
and, for any compatible pair $e,f \in E$, $S_e f = f$, $S_e U_f = U_f S_e = U_f$, $S_e S_f = S_f S_e$.
\\[0,5 cm]
Lemma 1: If $\mathbb{P}(f|e) = 1/2 = \mathbb{P}(f|e')$ for $e,f \in E$, then $S_e f = f'$ and $S_e f' = f$.
\\[0,5 cm]
Proof. $S_e f = 2 U_e f + 2 U_{e'} f - f = 2 \mathbb{P}(f|e) e + 2 \mathbb{P}(f|e') e' - f = e + e' - f = f'$.
The other identity follows by exchanging $f$ with $f'$. \hfill $\Box$
\\[0,5 cm]
Quantum teleportation and Grover's algorithm require some manipulations 
of the physical system under consideration. In the usual 
Hilbert space setting, these manipulations are represented by unitary transformations.
In the setting of this paper, the transformations $S_e$ shall 
take over their role. The transformations $S_e$ are linear and invertible, but generally they lack 
positivity and $S_e(f) \in A$ need not lie in $E$ for $f \in E$. 
This shall be resolved by the following assumption.
\\[0,5 cm]
\textbf{Assumption 1:}
$S_e E \subseteq E$ for each $e$ in $E$.
\\[0,5 cm]
Assumption 1 implies that each $S_e$ is a positive linear operator 
and thus becomes an automorphism of 
the order-unit space $A$. 
The restriction of $S_e$ to $E$ is
an automorphism of the quantum logic $E$ and $\mathbb{P}$
is invariant under $S_e$.

The positivity of the linear operators $S_e$ ($e \in E$) 
was already studied earlier; it is equivalent 
to a certain interesting property of the conditional 
probabilities restricting their second-order interference \cite{nie2012AMP} 
and has some further interesting consequences \cite{nie2014GenQTh}.
\\[0,5 cm]
\textbf{3.2 Sequential conditionalization}
\\[0,5 cm]
For a state $\rho$ and $e_1 \in E$ with $\rho(e_1) > 0$, 
the process of probability conditionalization 
can be repeated. The state $\rho_1$, defined by
$\rho_1(f) := \rho(f|e_1)$ for $f \in E$, can
be conditionalized a second time by $e_2 \in E$ 
with $\rho_1(e_2) = \rho(e_2|e_1)> 0$. The doubly 
conditionalized state $\rho_1(f|e_2)$
is denoted by $\rho(f|e_1, e_2)$. Then
$\rho(f|e_1, e_2) = \rho(U_{e_1}U_{e_2}f)/\rho(U_{e_1}e_2)$.

If this doubly conditionalized probability becomes
independent of the state $\rho$, it is again denoted by
$\mathbb{P}(f|e_1, e_2)$. Then 
\begin{center}
$\mathbb{P}(f|e_1, e_2) = p$
if and only if $U_{e_1}U_{e_2}f = p U_{e_1}e_2$.
\end{center}

In physical terms, the following assumption concerns a series of three sequential measurements,
where the first and the third measurement test the same property $e$, while 
the second measurement tests another property $f$ such that $\mathbb{P}(f|e)$ 
exists. The assumption states that, after the previous outcomes $e$ and $f$ 
in the first and second measurements, the probability for the outcome $e$ again in the 
third measurement shall be the same as the probability of the outcome $f$ 
in the second measurement after only the first measurement has been performed
and given the outcome $f$. It is hard to understand why nature should behave like this,
but it will later be seen that quantum mechanics satisfies this assumption. 
\\[0,5 cm]
\textbf{Assumption 2:} If $\mathbb{P}(f|e)$ exists for $e,f \in E$, then $\mathbb{P}(e|e,f)$ exists and $\mathbb{P}(e|e,f) = \mathbb{P}(f|e)$.
\\[0,5 cm]
In this case, it follows that: 
$\mathbb{P}(e'|e,f) = 1 -\mathbb{P}(f|e)$, 
$\mathbb{P}(e|e,f') = \mathbb{P}(f'|e) = 1 - \mathbb{P}(f|e)$, and
$\mathbb{P}(e'|e,f') = \mathbb{P}(f|e)$.

For two atoms $e$ and $f$, Assumption 1 implies that $\mathbb{P}(f|e) = \mathbb{P}(e|f)$. 
Note that $\mathbb{P}(a|e,f) = \mathbb{P}(a|f)$ holds for any $a$ and $e$, if $f$ is an atom.
This symmetry property is one of the so-called pure state properties, which Alfsen and Shultz 
use in their characterization of the state spaces of operator algebras \cite{AS02}.
Assumption 2 is a more general version of this property applicable also in cases when there are no atoms or pure states.
\\[0,5 cm]
Lemma 2: Suppose that Assumption 1 holds and that $\mathbb{P}(f|e)$ exists for $e,f \in E$. 
\begin{enumerate}
   \item[(a)]
   $U_e U_f e = (\mathbb{P}(f|e))^{2} e$,   $U_e U_{f'} e = (1 - \mathbb{P}(f|e))^{2} e$ and 
	 \newline
	 $U_e U_f e' = \mathbb{P}(f|e) (1 - \mathbb{P}(f|e)) e = U_e U_{f'} e'$.
   \item[(b)]
   $\mathbb{P}(S_fe|e) = (2 \mathbb{P}(f|e) -1)^{2}$
   \item[(c)]
   If $\mathbb{P}(f|e) = 1/2$, then $S_f e$ and $e$ are orthogonal.
\end{enumerate}
Proof. (a) $U_e U_f e = \mathbb{P}(e|e,f) U_f e = \mathbb{P}(e|e,f) \mathbb{P}(f|e) e = \mathbb{P}(f|e)^{2} e$.
The next identify follows from this one by exchanging $f$ with $f'$. Moreover,
$U_e U_f e' = U_e U_f \mathbb{I} - U_e U_f e = U_e f - \mathbb{P}(f|e)^{2} e = \mathbb{P}(f|e) e - \mathbb{P}(f|e)^{2}e $.
The last equality again follows by exchanging $f$ with $f'$.
\newline
(b) Define $p:= \mathbb{P}(f|e)$. Then, by (a), 
$U_e S_f e = 2 U_e U_f e + 2 U_e U_{f'}e - U_e e = 2 p^{2} e + 2 (1-p)^{2} e - e$,
and therefore $\mathbb{P}(S_fe|e) = 2 p^{2} + 2 (1-p)^{2} - 1 = (2p-1)^{2}$.
\newline
(c) By (b), $\mathbb{P}(S_fe|e) = 0$, which implies the orthogonality. \hfill $\Box$
\\[0,5 cm]
In the remaining part of this paper, the quantum logic $E$ shall always satisfy the Assumptions 1 and 2.
\newpage
\noindent
\large
\textbf{4. Entanglement-assisted quantum teleportation}
\normalsize
\\[0,5 cm]
The scenario for entanglement-assisted quantum teleportation consists
of two parties, named Alice and Bob, and three identical quantum systems with the labels $A,B,C$. The system
with label $C$ is in Alice's possession and she shall `teleport' its unknown system property to Bob
by sending some classical information to him.
The other two systems (labels $A$ and $B$) are initially `entangled' and the `entangled' property is known 
to both Alice and Bob. The system with label $B$ is given to Bob, the one with label $A$ to Alice.
She then performs a measurement on the combined system consisting of the two systems
with the labels $A$ and $C$. The outcome determines the classical information she sends to Bob.
He can then manipulate the system with label $B$ in such a way that it owns
the unknown initial property of the system with label $C$. This is consistent with the 
no-cloning theorem, because Alice's measurement on the combined system destroys the initial 
property of the system with label $C$.

For this scenario, consider the quantum logic $E$, a further quantum logic $E_o$, 
also possessing unique conditional probabilities, 
and two elements $e,f \in E_o$ 
with $1/2 = \mathbb{P}(f|e) = \mathbb{P}(f|e') = \mathbb{P}(e|f) = \mathbb{P}(e|f')$.
Then assume that $E$ contains three compatible copies of this quantum logic $E_o$. This means
that there are three morphisms 
$\pi_A : E_o \rightarrow E$, $\pi_B : E_o \rightarrow E$, $\pi_C : E_o \rightarrow E$ and that 
the subsets $\pi_A(E_o)$, $\pi_B(E_o)$, $\pi_C(E_o)$ of $E$ are pairwise compatible with each other.
The subset $\pi_C(E_o)$ represents the system in Alice's possession. The subsets $\pi_A(E_o)$ and $\pi_B(E_o)$
represent the other two initially `entangled' systems, the first one is given to Alice and the second one to Bob.

Furthermore suppose that there are two elements 
$d_{AB}$ and $d_{AC} \in E$ satisfying the following four conditions:
\begin{enumerate}
\item[(i)]
$ d_{AB} $ is compatible with $\pi_C E_o$, and $ d_{AC} $ is compatible with $\pi_B E_o$.
\item[(ii)]
$S_{\pi_A e} S_{\pi_B e} d_{AB} = d_{AB} = S_{\pi_A f} S_{\pi_B f} d_{AB}$.
\item[(iii)]
$1/2 = \mathbb{P}(\pi_A e|d_{AC}) = \mathbb{P}(\pi_A f|d_{AC})$ and
\newline
$ 1/2 = \mathbb{P} (\pi_A e \wedge \pi_C e | d_{AC}) 
= \mathbb{P} (\pi_A e' \wedge \pi_C e' | d_{AC})$.
\item[(iv)]
$\mathbb{P}(d_{AC} \wedge \pi_B x |d_{AB} \wedge \pi_C x) = \mathbb{P}(d_{AC} |d_{AB} \wedge \pi_C x) = 1/4$ for all $x$ in $E_o$.
\end{enumerate}
Now define $b_1 := d_{AC}$, $b_2 := S_{\pi_A e} d_{AC}$,  $b_3 := S_{\pi_A f} d_{AC}$ and $b_4 := S_{\pi_A e} S_{\pi_A f} d_{AC}$.
\\[0,5 cm]
Lemma 3: The elements $b_1, b_2, b_3, b_4$ of the quantum logic $E$ are pairwise orthogonal, and
$\mathbb{P}(b_k |d_{AB} \wedge \pi_C x) = 1/4 $ for all $x \in E_o $ and $k = 1,2,3,4$.
\\[0,5 cm]
Proof.
${b_1}$ and $b_2$ are orthogonal by (iii) and Lemma 2 (c). 
In the same way, ${b_3}$ and $b_4$ are orthogonal, since 
$$\mathbb{P}(\pi_A e|S_{\pi_A f}d_{AC})= \mathbb{P}(S_{\pi_A f} \pi_A e|d_{AC}) = \mathbb{P}(\pi_A e'|d_{AC}) = 1/2,$$
where the invariance of $\mathbb{P}(\ |\ )$ under morphisms 
has been used for the first equality and a further time for the
second equality to conclude that 
$S_{\pi_A f} \pi_A e = \pi_A e'$ from $\mathbb{P}(e|f) = \mathbb{P}(e|f') = 1/2$.

Furthermore, (iii) implies 
$\mathbb{P}((\pi_A e \wedge \pi_C e) \vee (\pi_A e' \wedge \pi_C e')|d_{AC}) = 1$
and therefore $b_1 = d_{AC} \leq (\pi_A e \wedge \pi_C e) \vee (\pi_A e' \wedge \pi_C e')$.
Then 
\begin{equation*}
\begin{split}
b_2 = S_{\pi_A e} b_1 &\leq S_{\pi_A e}((\pi_A e \wedge \pi_C e) \vee (\pi_A e' \wedge \pi_C e')) \\
&= (\pi_A e \wedge \pi_C e) \vee (\pi_A e' \wedge \pi_C e') \\
b_3 = S_{\pi_A f} b_1 &\leq S_{\pi_A f} ((\pi_A e \wedge \pi_C e) \vee (\pi_A e' \wedge \pi_C e')) \\
&= (S_{\pi_A f} \pi_A e \wedge S_{\pi_A f} \pi_C e) \vee (S_{\pi_A f} \pi_A e' \wedge S_{\pi_A f} \pi_C e') \\
&= (\pi_A e' \wedge \pi_C e) \vee (\pi_A e \wedge \pi_C e') \\
b_4 =  S_{\pi_A e} b_3 &\leq S_{\pi_A e} ((\pi_A e' \wedge \pi_C e) \vee (\pi_A e \wedge \pi_C e')) \\
&= (\pi_A e' \wedge \pi_C e) \vee (\pi_A e \wedge \pi_C e')
\end{split}
\end{equation*}
Thus $b_1$ and $b_2$ are orthogonal to $b_3$ and $b_4$, since these two pairs lie below 
different orthogonal elements of the quantum logic.

In the case $k=1$, $\mathbb{P}(b_k |d_{AB} \wedge \pi_C x) = 1/4 $ for $x \in E_o $ is part of (iv). 
The other cases then follow from this first one 
by using (ii) and the invariance of $\mathbb{P}$ 
under $S_{\pi_A e} S_{\pi_B e}$ and $S_{\pi_A f} S_{\pi_B f}$. 
For $k=2$ apply $S_{\pi_A e} S_{\pi_B e}$, for $k=3$ apply $S_{\pi_A f} S_{\pi_B f}$,
and for $k=4$ apply both one after the other.
In doing so, note that
the compatibility assumptions imply 
that $S$-transformations with different labels $A, B, C$ commute,
that $\pi_C x$ is invariant under $S_{\pi_A e}$, $S_{\pi_A f}$, $S_{\pi_B e}$ and $S_{\pi_B f}$
and that $d_{AC}$ is invariant under $S_{\pi_B e}$ and $S_{\pi_B f}$. \hfill $\Box$
\\[0,5 cm]
The element $d_{AB}$ in $E$ represents the entangled property of the combined system consisting of the two systems
with the labels $A$ and $B$, and $x$ represents the unknown property of the system with label $C$. Initially,
the combination of all three systems owns the property $d_{AB} \wedge \pi_C x$ in $E$.

Alice's measurement tests which one of the four orthogonal properties $b_1$, $b_2$, $b_3$ and $b_4$ 
the combined system under her control (labels $A$ and $C$) has; $b_1$, $b_2$, $b_3$ and $b_4$ each  
occur with the same probability $1/4$. If the outcome is $b_k$, the sequential conditional probability
with the first condition $d_{AB} \wedge \pi_C x$ and the second condition $b_k$ is to be determined.
\\[0,5 cm]
$k=1$: This case is a consequence of (iv) in the following way:
\begin{equation*}
\begin{split}
\mathbb{P}(\pi_B x|d_{AB} \wedge \pi_C x, b_1)  
&= \mathbb{P}(\pi_B x|d_{AB} \wedge \pi_C x, d_{AC})\\
&= \frac{\mathbb{P}(d_{AC} \wedge \pi_B x |d_{AB} \wedge \pi_C x)}{\mathbb{P}(d_{AC} |d_{AB} \wedge \pi_C x)} = 1
\end{split}
\end{equation*}
$k=2$: Apply $S_{\pi_A e} S_{\pi_B e}$ to the identity for $k=1$ 
and use the different invariances as in the proof of Lemma 3:
\begin{equation*}
\begin{split}
1 &= \mathbb{P}(S_{\pi_A e} S_{\pi_B e} \pi_B x| S_{\pi_A e} S_{\pi_B e} d_{AB} \wedge S_{\pi_A e} S_{\pi_B e} \pi_C x, S_{\pi_A e} S_{\pi_B e} b_1) \\
&= \mathbb{P}(S_{\pi_B e} S_{\pi_A e} \pi_B x| d_{AB} \wedge \pi_C x, S_{\pi_A e} b_1) \\
&= \mathbb{P}(S_{\pi_B e} \pi_B x| d_{AB} \wedge \pi_C x, b_2)
\end{split}
\end{equation*}
$k=3$: Apply $S_{\pi_A f} S_{\pi_B f}$ and proceed in the same way as in the last case:
$$1 = \mathbb{P}(S_{\pi_B f} \pi_B x| d_{AB} \wedge \pi_C x, b_3)$$
$k=4$: Apply both $S_{\pi_A f} S_{\pi_B f}$ and $S_{\pi_A e} S_{\pi_B e}$  
one after the other and proceed in the same way as in the last two cases:
$$1 = \mathbb{P}(S_{\pi_B e} S_{\pi_B f} \pi_B x| d_{AB} \wedge \pi_C x, b_4)$$
Alice communicates to Bob, which one of the four cases $b_k$, $k=1,2,3,4$, 
is her measurement outcome; two classical bits are sufficient for this communication.
The sequential conditional probability, calculated above, shows that, 
in the case $k=1$, Bob's system (label B) 
now has the property $\pi_B x$ with probability $1$. 
This means that the initial unknown property of the 
system with label $C$ was successfully transferred to the system with label $B$.
In the other cases, Bob knows how to manipulate his system in order to achieve
that it has the property $\pi_B x$: he performs the transformations 
$S_{\pi_B e}$ in the case $k=2$,
$S_{\pi_B f}$ in the case $k=3$, and
$S_{\pi_B e} S_{\pi_B f}$ in the case $k=4$.
Note that all these transformations are their own inverse.

The link between this abstract setting 
and usual quantum teleportation may not be immediately visible,
but will be revealed later in section 6, where 
Hilbert space quantum mechanics is considered.
\\[0,5 cm]
\large
\textbf{5. Grover's quantum search algorithm}
\normalsize
\\[0,5 cm]
\textbf{5.1 A further assumption}
\\[0,5 cm]
A further assumption, which is needed for the treatment 
of Grover's algorithm in the following subsection,  
shall be introduced first.
Again it is hard to understand why nature should
behave like this, but it will later be seen 
that quantum mechanics satisfies this assumption.
\\[0,5 cm]
\textbf{Assumption 3:} For the states $\rho$ and elements $f$ in $E$ with
$\rho(f) = 0$, the identity $\rho(f|e) \rho(e) = \rho(f|e') \rho(e')$ 
shall hold for all $e \in E$.
\\[0,5 cm]
Lemma 4: Assumption 3 is equivalent to the following condition:
$U_{f'} U_e f = U_{f'} U_{e'} f$ for all $e,f$ in $E$.
\\[0,5 cm]
Proof: For $e,f \in E$ and a state $\rho$ with $\rho(f) \neq 1$, first define 
$\rho_{f'}$ by $\rho_{f'}(a):= \rho (U_{f'} a) / \rho(f')$ for $a \in E$. 
Then $\rho_{f'} (f) = 0$ and Assumption 3 yields
$\rho_{f'}(f|e) \rho_{f'}(e)$ $=$ $\rho_{f'}(f|e') \rho_{f'}(e')$. Thus 
$ \rho (U_{f'} U_e f) = \rho (U_{f'} U_{e'} f)$. This identity holds
for all states $\rho(f)$ and also when $\rho(f) = 1$, 
since both sides then equal $0$. Therefore 
$U_{f'} U_e f = U_{f'} U_{e'} f$.

Vice versa, assume $U_{f'} U_e f = U_{f'} U_{e'} f$ and $\rho(f) = 0$ with a state $\rho$ and $e,f$ in $E$. 
Then $\rho(f') = 1$ and $\rho = \rho U_{f'}$. Therefore
$\rho(f|e) \rho(e) = \rho U_e f = \rho U_{f'} U_e f = \rho U_{f'} U_{e'} f = \rho U_{e'} f = \rho(f|e') \rho(e')$. \hfill $\Box$
\\[0,5 cm]
In the remaining part of section 5, the quantum logic $E$ shall now satisfy 
Assumption 3 in addition to Assumptions 1 and 2.
\\[0,5 cm]
\textbf{5.2 The algorithm}
\\[0,5 cm]
Suppose that an unsorted data base with $n$ indexed entries 
contains one specific entry 
satisfying a certain search criterion. 
The task of the algorithm is to find the index of this entry.
Assume that this index is $k_o$.

Now consider $n$ pairwise orthogonal elements $f_k$ in the quantum logic $E$
and a further element $e \in E$ with $\mathbb{P}(f_k|e) = 1/n = \mathbb{P}(e|f_k)$ for $k = 1,2,...,n$.
The initial property of the system is $e$. The system shall then be manipulated
in such a way that the probability of getting the outcome $f_{k_o}$
in a measurement of the $f_k$, $k=1,2,...,n$, becomes close to 1.

This manipulation is a repeated application 
of the transformations $S_{f_{k_o}}$ and $S_e$. After
the $r$-th iteration step, the initial property has been transformed to 
$(S_e S_{f_{k_o}})^{r}e$, and $\mathbb{P}(f_k|S_e S_{f_{k_o}})^{r}e)$ is the probability
of getting the outcome $f_k$ in the measurement after the $r$-th iteration step.
For the desired outcome $f_{k_o}$, the probability becomes  
$$\mathbb{P}\left(f_{k_o} | (S_e S_{f_{k_o}})^{r}e \right) = sin^{2}\left(\left(2r+1\right)arcsin \left( \frac{1}{\sqrt{n}} \right) \right)$$ 
by Lemma 5 in the annex.
This is exactly the well-known success probability of Grover's algorithm in the 
usual quantum mechanical realm, which has been reproduced here in a 
much more abstract and general setting and under more general assumptions. 
It is interesting to note that it becomes 1 in the case $n=4$ and $r=1$; this means that,
if the data base consists of four entries only, 
the algorithm outputs the correct result after the first step already 
and with $100 \%$ certainty. In general, however, the algorithm is 
not deterministic.
The required number of iterations resulting from the above formula
and the speed-up versus classical search algorithms are well-known.  
For further information, it is referred to the extensive literature
concerning Grover's algorithm. 

Notwithstanding the differences between the two approaches and between the used physical principles, 
the above result is in line with recent work by C. M. Lee and J. H. Selby \cite{Lee-Selby2016}
who found out that, concerning the search algorithm, post-quantum interference
does not imply a computational speed-up over quantum theory.
Post-quantum interference means interference of third or higher order 
in Sorkin's hierarchy \cite{sorkin1994quantum}
and represents an interesting potential property of
the conditional probabilities \cite{nie2012AMP},
but the above result holds for interference of second order (quantum interference)
as well as for all higher orders and is independent of the actual order.
However, Lee and Selby study only a bound for the computational speed-up;
they leave open the question whether or when an algorithm exists
that achieves this bound.

In the following section, 
Grover's algorithm will be reconsidered 
to elucidate the link between the above
version and its usual Hilbert space version.
\newpage
\noindent
\large
\textbf{6. Usual quantum mechanics}
\normalsize
\\[0,5 cm]
\textbf{6.1 The Hilbert quantum logic}
\\[0,5 cm]
Quantum mechanics uses a special quantum logic; 
it consists of the self-adjoint projection operators $e$
(i.e., $e = e^{*}$ and $e = e^{2}$) on a 
Hilbert space $H$ and is an orthomodular lattice. 
The identity operator becomes the element 
$\mathbb{I}$ of the quantum logic. Compatibility 
here means that the self-adjoint projection operators
commute. The unique conditional probabilities exist;
it has been shown in Ref. \cite{niestegge2001non} that,
with two self-adjoint projection operators $e$ and $f$ on $H$, the
conditional probability has the shape
$$\rho(f|e) = \frac{trace(aefe)}{trace(ae)} = \frac{trace(eaef)}{trace(ae)}$$
for a state $\rho$ defined by the statistical operator $a$ (i.e., $a$
is a self-adjoint operator on $H$ with non-negative spectrum and $trace(a)=1$).
This means that $U_e y = eye$ for operators $y$ on $H$. Here
$ae$, $efe$, $eye$, and so on, denote the usual operator product 
of the operators $a, e, f, y$.

The above identity reveals
that conditionalization becomes identical with the state transition of the L\"uders
- von Neumann measurement process. Therefore, the conditional probabilities
can be regarded as a generalized mathematical model of projective quantum
measurement.

$\mathbb{P}(f|e)$ exists with $\mathbb{P}(f|e) = p$ if and only if the operators
$e$ and $f$ on $H$ satisfy the algebraic identity $efe = pe$.
This transition probability between the outcomes of two consecutive measurements
is independent of any underlying state. The algebraic identity $efe = pe$
clearly demonstrates that the probability $p = \mathbb{P}(f|e)$ 
results solely from the algebraic structure
of the quantum logic.

The atoms are the self-adjoint projections on the one-dimensional subspaces 
of $H$; if $e$ is an atom and $| \xi \rangle $ a normalized vector in the 
corresponding one-dimensional subspace, then 
$\mathbb{P}(f|e) = \left\langle \xi| f \xi\right\rangle$.
The atomic states thus coincide with the quantum mechanical pure states or vector states,
which are often called wavefunctions.
If $f$ is an atom, too, and $| \eta \rangle $ a normalized vector in the 
corresponding one-dimensional subspace, then 
$\mathbb{P}(f|e) = \left|\left\langle \eta| \xi\right\rangle\right|^{2}$.
\\[0,5 cm]
\textbf{6.2 Assumptions 1, 2 and 3 revisited}
\\[0,5 cm]
It shall now be checked whether the Hilbert space quantum logic
satisfies the assumptions 1, 2 and 3. For two elements $e$ and $f$
in this quantum logic, 
\begin{align*}
	S_e f &= 2 U_e f + 2 U_{e'} f - f 
= 2 efe + 2 (\mathbb{I} - e)f(\mathbb{I}-e) - f \\
&= (2e -\mathbb{I})f(2e -\mathbb{I})            \\
&= (e - e')f(e - e').
\end{align*}
The operator $2e -\mathbb{I} = e - e'$ is unitary. Therefore $S_e f$ is 
a self-adjoint projection operator on $H$ as $f$ is. This means
that $S_e f$ belongs to the quantum logic, whenever $f$ does,
and Assumption 1 is satisfied.

Note that $S_e f = (e-e')f(e-e') = (e'-e)f(e'-e)$, though 
the operators $e - e'$ and $e' - e = - (e-e')$ act differently on the 
Hilbert space elements. The effect are different signs.
However, it is well-known that not the individual 
Hilbert space element, but the ray or one-dimensional linear
subspace it generates is relevant in quantum mechanics. This 
ray or subspace is not affected  by the sign change.

Now suppose $\mathbb{P}(f|e)=p$. Then $efe = U_e f = p e$
and $U_e U_f e = efefe = (efe)(efe)= (pe)(efe) = pefe = p U_e f$. 
This means $\mathbb{P}(e|e,f) = p = \mathbb{P}(f|e)$, and 
Assumption 2 is satisfied.

Assumption 3 is checked using Lemma 4. 
$U_{f'} U_{e'} f = f'(\mathbb{I} - e)f(\mathbb{I} - e)f' = 
f'ff' - f'eff' - f'fef' + f'efef' = f'efef' = U_{f'} U_{e} f$.
Thus Assumption 3 is satisfied as well.

Assumption 1 is of a mathematical technical type, but the other
two assumptions represent very interesting properties of the
quantum mechanical probabilities, though it is hard to understand
why nature should possess these special properties. The
property forming Assumption 3 has 
been detected by T. Fritz \cite{Fritz2010JMP}. 
The property forming Assumption 2 appears 
for the first time in this paper.

The results of sections 4 and 5 shall now be applied to
the special Hilbert space quantum logic
in order to reveal the link to the well-known
Hilbert space versions of quantum teleportation
and Grover's algorithm. It is started with 
Grover's algorithm. 
\\[0,5 cm]
\textbf{6.3 Grover's algorithm revisited}
\\[0,5 cm]
Again assume that
$k_o$ is the index of the data base entry it is searched for
among $n$ entries in total.
Let $ | k \rangle $, $k=1,...,n$, be $n$ pairwise 
orthogonal normalized elements of the Hilbert space $H$.
Define 
\begin{center}
$\psi := \frac{1}{\sqrt{n}} \sum^{n}_{k=1} | k \rangle,$ 
$f_k := | k \rangle \langle k | $
and $e := | \psi \rangle \langle \psi |$. 
\end{center}
These are elements of the Hilbert space quantum logic and 
they satisfy 
$\mathbb{P}(f_k|e) = \mathbb{P}(e|f_k) = |\left\langle k| \psi \right\rangle|^{2} = 1/n$.
As seen in 6.2, $S_e$ and $S_{f_{k_o}}$ can be represented
by the unitary operators $u_e := 2 e - \mathbb{I} = e - e'$ and 
$u_{k_o} := 2 f_{k_o} - \mathbb{I} = f_{k_o} - f_{k_o}'$. 
These are the operators
used in the Hilbert space version 
of Grover's algorithm; the first one is 
the so-called Grover diffusion operator. Then
$$(S_e S_{f_{k_o}})^{r}e = | (u_e u_{k_o})^{r}\psi \rangle \langle (u_e u_{k_o})^{r}\psi |$$
and the success probability of finding $k_o$ with a 
measurement after the $r$-th iteration step becomes
$$ |\langle k_o | (u_e u_{k_o})^{r}\psi \rangle |^{2} 
= \mathbb{P}\left(f_{k_o}|(S_e S_{f_{k_o}})^{r}e\right) 
= sin^{2}\left((2r+1)arcsin\left(\frac{1}{\sqrt{n}}\right)\right).$$ 
For the direct proof of this result in the Hilbert space setting,
using the unitary operators $u_e$ and $u_{k_o}$,
it is sufficient to consider a $2 \times 2 $-matrix. The proof
of the general Lemma 5 in the annex is more difficult, since  
the Jordan form of a $4 \times 4 $-matrix must be calculated. 

Note that the version of Grover's algorithm in 5.2
does not require that the $f_k$ and $e$ are atoms 
(i.e., projections on one-dimensional subspaces)
and thus becomes more general than the known 
version - even in usual quantum mechanics.
\\[0,5 cm]
\textbf{6.4 Teleportation revisited}
\\[0,5 cm]
The usual setting of entanglement-assisted 
quantum teleportation consists of three
two-dimensional Hilbert spaces $H_A$, $H_B$, $H_C$
and their tensor product $H_A \otimes H_B \otimes H_C$.
Each one of the two-dimensional Hilbert spaces 
has a basis denoted by $| 0 \rangle$ and $| 1 \rangle$
with the appropriate labels. 
Moreover, consider $ | \varphi \rangle:= \frac{1}{\sqrt{2}}(| 0 \rangle + | 1 \rangle) $ with the appropriate labels
in each one of the three Hilbert spaces $H_A$, $H_B$, $H_C$.
Furthermore,
the following two Hilbert space elements 
play an important role:
\\
\\
\hspace*{2 cm}
$| \psi_{AB} \rangle = \frac{1}{\sqrt{2}}(| 0_A \rangle \otimes | 0_B \rangle + | 1_A \rangle \otimes | 1_B \rangle) \in H_A \otimes H_B$ and
\\
\\
\hspace*{2 cm}
$| \psi_{AC} \rangle = \frac{1}{\sqrt{2}}(| 0_A \rangle \otimes | 0_C \rangle + | 1_A \rangle \otimes | 1_C \rangle) \in H_A \otimes H_C$.
\\

The mapping to the situation of section 4 works as follows. 
The quantum logic of the two-dimensional Hilbert space becomes $E_o$,
the quantum logic of the tensor product $H_A \otimes H_B \otimes H_C$
becomes $E$, and the morphisms $\pi_A, \pi_B, \pi_C$ map any $y$ in $E_o$ to 
$y_A \otimes \mathbb{I}_B \otimes \mathbb{I}_C$,
$\mathbb{I}_A \otimes y_B \otimes \mathbb{I}_C$,
$\mathbb{I}_A \otimes \mathbb{I}_B \otimes y_C$,
respectively. Define
$e :=| 1 \rangle \langle 1 |$ and $f :=| \varphi \rangle \langle \varphi |$
with the appropriate labels $A, B, C$ for each one 
of the three Hilbert space $H_A$, $H_B$, $H_C$.
Furthermore, define $d_{AB} := (| \psi_{AB} \rangle \langle \psi_{AB} |) \otimes \mathbb{I}_C $ and
$d_{AC} := | (\psi_{AC} \rangle \langle \psi_{AC} |) \otimes \mathbb{I}_B$
in the quantum logic of $H_A \otimes H_B \otimes H_C$.

Now the conditions (i) - (iv) in section 4 
shall be checked. The first one is satisfied, since
$d_{AB}$ commutes with all operators on $H_C$ and $d_{AC}$ 
commutes with all operators on $H_B$. Moreover,
$S_{e_A \otimes \mathbb{I}_B \otimes \mathbb{I}_B} S_{\mathbb{I}_A \otimes e_B \otimes \mathbb{I}_B} d_{AB} = d_{AB}$, 
since the unitary operators 
$(e_A - e_A') \otimes \mathbb{I}_B $ and $\mathbb{I}_A \otimes (e_B - e_B')$ 
only change the sign of 
$ | 0_{A} \rangle $ and $ | 0_{B} \rangle $ in  
$ | \psi_{AB} \rangle $, but together leave 
$ | \psi_{AB} \rangle $ invariant;
$S_{f_A \otimes \mathbb{I}_B \otimes \mathbb{I}_B} S_{\mathbb{I}_A \otimes f_B \otimes \mathbb{I}_B} d_{AB} = d_{AB}$, 
since the unitary operators 
$(f_A - f_A') \otimes \mathbb{I}_B $ and $ \mathbb{I}_A \otimes (f_B - f_B')$ 
exchange 
$ | 0_{A} \rangle $ with $ | 1_{A} \rangle $ and 
$ | 0_{B} \rangle $ with $ | 1_{B} \rangle $ in  
$ | \psi_{AB} \rangle $ and thus leave 
$ | \psi_{AB} \rangle $ invariant. Therefore,
(ii) is satisfied as well. Furthermore, 
\begin{align*}
\mathbb{P}(f_A \otimes \mathbb{I}_B \otimes \mathbb{I}_C|d_{AC}) 
&= \langle \psi_{AC} | (| \varphi_A \rangle \langle \varphi_A |) \otimes \mathbb{I}_C | \psi_{AC} \rangle  \\
&= \frac{1}{2} \langle \psi_{AC} | (| 0_A \rangle + | 1_A \rangle)(\langle 0_A | + \langle 1_A|) \otimes \mathbb{I}_C | \psi_{AC} \rangle = \frac{1}{2}.
\end{align*}
This is one of the identities of condition (iii), and similar calculations yield 
the other ones. Note that 
$(y_A \otimes \mathbb{I}_B \otimes \mathbb{I}_C) \wedge (\mathbb{I}_A \otimes y_B \otimes \mathbb{I}_C) \wedge (\mathbb{I}_A \otimes \mathbb{I}_B \otimes y_C)$ 
is the same as $y_A \otimes y_B \otimes y_C$ here
for any elements $y_A$, $y_B$, $y_C$ in the quantum logics 
of the Hilbert spaces $H_A$, $H_B$, $H_C$.
\newpage

Concerning (iv), consider 
$| \xi_B \rangle = \alpha | 0_B \rangle + \beta | 1_B \rangle$,
$| \xi_C \rangle = \alpha | 0_C \rangle + \beta | 1_C \rangle$, 
with some complex numbers $\alpha, \beta$ such that $|\alpha|^{2} + |\beta|^{2} = 1$, and  
$x_B = | \xi_B \rangle \langle \xi_B |$, $x_C = | \xi_C \rangle \langle \xi_C |$.
Then
$$\mathbb{P}(d_{AC} \wedge x_B | d_{AB} \wedge x_C) 
= \left| \langle \psi_{AC}, \xi_B | \psi_{AB}, \xi_C \rangle \right|^{2} 
= \left| \frac{1}{2} (\bar{\alpha} \alpha + \bar{\beta} \beta) \right|^{2} 
= \frac{1}{4}.$$
Furthermore, the orthogonal complement of $x_B$ has the shape
$x_B' = | \xi_B' \rangle \langle \xi_B' |$ with $ \xi_B' = \alpha' | 0_B \rangle + \beta' | 1_B \rangle$,
$|\alpha'|^{2} + |\beta'|^{2} = 1$ 
and $\bar{\alpha'} \alpha + \bar{\beta'} \beta = 0.$ Then
$$\mathbb{P}(d_{AC} \wedge x_B' | d_{AB} \wedge x_C) 
= \left| \langle \psi_{AC}, \xi_B' | \psi_{AB}, \xi_C \rangle \right|^{2} 
= \left| \frac{1}{2} (\bar{\alpha'} \alpha + \bar{\beta'} \beta) \right|^{2} 
= 0$$
and
$$ \mathbb{P}(d_{AC} | d_{AB} \wedge x_C) =  \frac{1}{4} + \mathbb{P}(d_{AC} \wedge x_B' | d_{AB} \wedge x_C) = \frac{1}{4}.$$

Alice's measurement tests which one of the four
properties $b_1$, $b_2$, $b_3$, $b_4$ the combined system 
under her control (labels $A$ and $C$) has. The first one is $b_1 = d_{AC}$;
this is the projection on the one-dimensional subspace of $H_A \otimes H_C$,
which is generated by $|\psi_{AC} \rangle$. The other ones 
are projections on the one-dimensional subspaces of $H_A \otimes H_C$
which are generated by:
\begin{align*}
((e_A - e_A')\otimes \mathbb{I}_C)|\psi_{AC}\rangle &= \frac{1}{\sqrt{2}}( | 1_A \rangle \otimes | 1_C \rangle - | 0_A \rangle \otimes | 0_C \rangle )            \\
((f_A - f_A')\otimes \mathbb{I}_C)|\psi_{AC}\rangle &= \frac{1}{\sqrt{2}}( | 1_A \rangle \otimes | 0_C \rangle + | 0_A \rangle \otimes | 1_C \rangle )            \\
((e_A - e_A')(f_A - f_A')\otimes \mathbb{I}_C)|\psi_{AC}\rangle &= \frac{1}{\sqrt{2}}( | 1_A \rangle \otimes | 0_C \rangle - | 0_A \rangle \otimes | 1_C \rangle)
\end{align*}
These four elements form a so-called Bell basis of $H_A \otimes H_C$, 
which is used by Alice in her local measurement in the usual Hilbert 
space treatment of quantum teleportation.
Depending on which one of these outcomes Alice's measurement provides
and according to section 4, 
Bob uses one of the following unitary transformations on $H_B$: 
\\
\\
\hspace*{3,25 cm}
 - the identity in the first case,
\\
\hspace*{3,25 cm}
 - $e_B - e_B'$ in the second case,
\\
\hspace*{3,25 cm}
 - $f_B - f_B'$ in the third case, and 
\\
\hspace*{3,25 cm}
 - $(e_B - e_B')(f_B - f_B')$ in the last case.
\\
\\
These four transformations coincide with the unitary operations 
occurring as Bob's operations in the usual Hilbert 
space treatment of quantum teleportation.
After the transformation, the initial
property of the system with the label $C$
has successfully been transferred to Bob's system (label B). 

Note that the version of quantum teleportation
considered in section 4 does not require that 
$e$ and $f$ are atoms 
(i.e., projections on one-dimensional subspaces)
and does not need the tensor product. It thus 
becomes more general than the known 
version - even in usual quantum mechanics - just as 
the version of Grover's algorithm
considered in section 5.
\\[0,5 cm]
\large
\textbf{7. Conclusion}
\\[0,5 cm]
\normalsize
Some major features of quantum information theory are
the no-cloning theorem, quantum key distribution,
entanglement-assisted quantum teleportation 
and Grover's search algorithm. 
In this paper and two earlier ones, these features
have been transferred to a general and abstract 
setting - a non-classical extension of conditional probability -,
which shows that they do not necessarily require
the usual Hilbert space quantum mechanics, but 
allow a much more abstract access and
exist in a much more general theory.
Equally important may be that, even in usual quantum mechanics,
more cases are covered, since any system properties
and not only the atomic ones (or pure states)
can be used.

The question now suggests itself,
whether Shor's factoring algorithm \cite{shor1994} - a further 
important result in quantum information theory -
can also be transferred to the same setting.
The transformations $S_e$ are available in this general setting
and are sufficient for quantum teleportation and Grover's algorithm.
Shor's factoring algorithm, however, seems to need more. It uses
the so-called quantum Fourier transformation 
which requires the complex numbers and 
appears to be available only 
in the complex Hilbert space or von Neumann algebras. 
It does not seem to be possible 
to gain such a transformation in the general setting. 
A positive answer to the question above is therefore not
expected.

Another famous quantum algorithm is due to D. Deutsch and R. Jozsa 
\cite{Cleve-Ekert-et-al1998, Deutsch1985, Deutsch-Jozsa1992}.
Although fundamental obstacles are not immediately obvious, it is currently
not clear whether it can be transferred 
to the general and abstract setting
in the same was as Grover's algorithm. A first barrier is
the implementation of the so-called quantum
oracle needed here.

Along the way, an interesting new property of quantum mechanics
(Assumption 2 in sections 3 and 6) has been detected in this paper. 
It concerns the sequential conditionalization or,
in physical terms, three sequential measurements, where the first
and third measurement test the same system property 
while a different incompatible property is tested in between 
in the second measurement. Under certain conditions,
the probabilities for the outcomes in the second and third
measurement must then be identical, although different and incompatible
system properties are measured.

\bibliographystyle{abbrv}
\bibliography{Literatur}

\newpage
\noindent
ANNEX
\\[0,5 cm]
Lemma 5: Suppose that the quantum logic $E$ satisfies the Assumptions 1, 2 and 3
and that $\mathbb{P}(f|e) = p = \mathbb{P}(e|f)$ for some $e,f \in E$. Then, for $r = 1, 2, 3, ...$,
$$\mathbb{P}\Bigl(f|(S_e S_f)^{r}e\Bigr) = \mathbb{P}\Bigl((S_f S_e)^{r}f|e\Bigr) = sin^{2}\Bigl((2r+1)arcsin(\sqrt{p})\Bigr).$$
Proof. 
The first equality follows from the invariance of $\mathbb{P}(\ |\ )$ under the automorphism $(S_f S_e)^{r}$,
which is its own inverse. For the proof of the second equality,
consider the following four elements in the order-unit space $A$: 
$b_1 := e$, $b_2 := f$, $b_3 := U_{e'} f$ and $b_4 := U_{f'} e$.
Note that Lemmas 2 (a) and 4 are repeatedly applied in the following calculations.
\\
\\
$S_f S_e b_1 = S_f S_e e = S_f e = 2 U_f e + 2 U_{f'} e - e = 2 p f + 2 U_{f'} e - e = - b_1 + 2 p b_2 + 2 b_4$
\\
\\
Then use the identity $S_e f = 2 U_e f + 2 U_{e'} f - f = 2 p e + 2 U_{e'} f - f$ to get
\\
\\
$ S_f S_e b_2 = S_f S_e f = 2 p S_f e + 2 S_f U_{e'} f - S_f f $
\\[0,1 cm]
$= 2 p (2 U_f e + 2 U_{f'} e - e) + 2 (2 U_f U_{e'} f + 2 U_{f'} U_{e'} f - U_{e'} f) - f $
\\[0,1 cm]
$= 2 p (2 p f + 2 U_{f'} e - e) + 2 (2 (1-p)^{2} f + 2 U_{f'} U_{e} f - U_{e'} f) - f $
\\[0,1 cm]
$= 2 p (2 p f + 2 U_{f'} e - e) + 2 (2 (1-p)^{2} f + 2 p U_{f'} e - U_{e'} f) - f $
\\[0,1 cm]
$= - 2 p e + (8p^{2} - 8p + 3) f - 2 U_{e'} f + 8 p U_{f'} e $
\\[0,1 cm]
$= - 2 p b_1 + (8p^{2} - 8p + 3) b_2 - 2 b_3 + 8 p b_4 $
\\
\\
$S_f S_e b_3 = S_f S_e  U_{e'} f = S_f U_{e'} f = 2 U_f U_{e'} f + 2 U_{f'} U_{e'} f - U_{e'} f $
\\[0,1 cm]
$= 2 (1-p)^{2} f + 2 U_{f'} U_{e} f - U_{e'} f $
$= 2 (1-p)^{2} f + 2 p U_{f'} e - U_{e'} f $
\\[0,1 cm]
$= 2 (1-p)^{2} b_2 - b_3 + 2 p b_4$
\\
\\
$S_f S_e b_4 = S_f S_e U_{f'} e = 2 S_f U_e U_{f'} e + 2 S_f U_{e'} U_{f'} e - S_f U_{f'} e $
\\[0,1 cm]
$= 2 (1 - p)^{2} S_f e + 2 S_f U_{e'} U_{f} e - U_{f'} e = 2 (1 - p)^{2} S_f e + 2 p S_f U_{e'} f - U_{f'} e $
\\[0,1 cm]
$= 2 (1 - p)^{2} (2 U_f e + 2 U_{f'}e - e) + 2 p ( 2 U_f U_{e'} f + 2 U_{f'} U_{e'} f - U_{e'} f ) - U_{f'} e $
\\[0,1 cm]
$= 2 (1 - p)^{2} (2 p f + 2 U_{f'}e - e) + 2 p (2(1-p)^{2} f + 2 U_{f'} U_{e} f - U_{e'} f) - U_{f'} e $
\\[0,1 cm]
$= 2 (1 - p)^{2} (2 p f + 2 U_{f'}e - e) + 2 p (2(1-p)^{2} f + 2 p U_{f'} e - U_{e'} f) - U_{f'} e $
\\[0,1 cm]
$= - 2 (1-p)^{2} e + 8 p (1-p)^{2} f  - 2 p U_{e'} f + (8 p^{2} - 8 p + 3) U_{f'} e $
\\[0,1 cm]
$= - 2 (1-p)^{2} b_1 + 8 p (1-p)^{2} b_2  - 2 p b_3 + (8 p^{2} - 8 p + 3) b_4 $
\\
\\
The linear subspace in $A$, generated by $b_1, b_2, b_3, b_4$, 
is invariant under $S_f S_e$, which follows from the above identities.
With respect to this basis, the restriction of $S_f S_e$ to this subspace is represented 
by the following matrix:
\newpage
$$ M := 
	\begin{pmatrix}
-1 & - 2 p & 0 & -2 (1-p)^{2}\\
\\
2 p & 8p^{2} - 8p + 3  & 2 (1-p)^{2} & 8 p(1-p)^{2} \\
\\
0 & -2 & -1 & -2p \\
\\
2 & 8p & 2p & 8p^{2} - 8p + 3
 \end{pmatrix}
$$
The Jordan form of this $4\times4$matrix is now computed in two steps,
each one basically dealing with the better manageable $2\times2$-matrices. 
First consider the following matrix $N_1$ 
$$ N_1 = 
	\begin{pmatrix}
          1-p      &        0        &    1-p       &    0   \\
\\
           0       &       1-p       &    0         &   1-p     \\
\\
           -1      &        0        &    1         &    0   \\
\\
            0      &        -1       &    0         &    1   \\
 \end{pmatrix} $$
and its inverse 
$$ N_1^{-1} = \frac{1}{2(1-p)}
	\begin{pmatrix}
             1   \  &     \   0   \   &       p-1       &     0   \\
\\
             0      &        1        &         0       &     p-1 \\
\\
             1      &         0       &       1-p       &     0   \\
\\
             0      &         1       &         0       &    1-p  \\
 \end{pmatrix} $$
Then 
$$N_1^{-1}MN_1 = 
	\begin{pmatrix}
            -1          &   2-4p         &       0      &    0   \\
\\
         -2 + 4p        &  (3-4p)(1-4p)  &       0      &    0   \\
\\
             0          &    0           &       -1     &   -2   \\
\\
             0          &    0           &       2      &   3    \\
 \end{pmatrix}
$$
The Jordan forms of the two $2 \times 2$ submatrices 
top left and bottom right can be calculated separately. With
$$
N_2 = 
	\begin{pmatrix}
            1                       &                 1                   &       0      &    0   \\
\\
   1 - 2p + 2 \sqrt{p(1-p)} \ i       &       1 - 2p - 2 \sqrt{p(1-p)} \ i    &       0      &    0   \\
\\
            0                       &                 0                   &       1      &    -2 \\
\\
            0                       &                 0                   &       0      &     2
 \end{pmatrix}
$$
and
$$
N_2^{-1} = 
	\begin{pmatrix}
   \frac{1}{2} + \frac{1-2p}{4 \sqrt{p(1-p)}} i     &  \frac{-1}{4\sqrt{p(1-p)}} i  &       0      &    0   \\
\\
   \frac{1}{2} - \frac{1-2p}{4 \sqrt{p(1-p)}} i     &  \frac{1}{4\sqrt{p(1-p)}} i   &       0      &    0   \\
\\
                    0                 &              0                       &       1      &    1    \\
\\
                    0                 &              0                       &       0      &  \frac{1}{2} \\
 \end{pmatrix}
$$
the desired Jordan form of $M$ is: 
$$ N_2^{-1} N_1^{-1} M N_1 N_2 = 
	\begin{pmatrix}
     \alpha_1   &     0     &       0      &    0   \\
\\
       0       &  \alpha_2  &       0      &    0   \\
\\
       0       &     0      &       1      &    0   \\
\\
       0       &     0      &       1      &    1 \\
 \end{pmatrix}
$$
where
$$\alpha_1 = 8p^{2} - 8p + 1 + 4(1-2p) \sqrt{p(1-p)} \ i, $$  
$$\alpha_2 = 8p^{2} - 8p + 1 - 4(1-2p) \sqrt{p(1-p)} \ i  $$ 
and $1$ are the eigenvalues of $M$. 
This (almost diagonal) matrix can now easily be raised to the $r$-th power, and $M^{r}$ can be calculated:
$$M^{r} = N_1 N_2
	\begin{pmatrix}
     \alpha_1^{r}   &     0     &       0      &    0   \\
\\
       0       &  \alpha_2^{r}  &       0      &    0   \\
\\
       0       &     0          &       1      &    0   \\
\\
       0       &     0          &       r      &    1   \\
 \end{pmatrix}
N_2^{-1} N_1^{-1}
$$
$$ = 
	\begin{pmatrix}
      \cdots       &    - r + \frac{1}{4\sqrt{p(1-p)}} Im (\alpha_1^{r})                                              &      \cdots        &    \cdots      \\
			                                                                                                                                                   \\
      \cdots       &     \frac{1}{2}(2 r + 1 + Re(\alpha_1^{r}) + \frac{1-2p}{2\sqrt{p(1-p)}} Im(\alpha_1^{r}))        &      \cdots        &    \cdots      \\
			                                                                                                                                                   \\
      \cdots       &     \cdots                                                                                    &      \cdots        &    \cdots      \\
			                                                                                                                                                   \\ 
			\cdots       &     \frac{1}{2(1-p)} (2 r + 1 - Re(\alpha_1^{r}) - \frac{1-2p}{2\sqrt{p(1-p)}} Im(\alpha_1^{r}))  &      \cdots        &    \cdots      \\
 \end{pmatrix}
$$
Here, $Re(z)$ [$Im(z)$] denotes the real [imaginary] part of the complex number $z$.
Since $\alpha_2$ is the complex conjugate of $\alpha_1$, it does not anymore appear in this matrix. 
Note that only the second column is displayed, 
since only these entries will be used for the following calculation 
of $U_e (S_f S_e)^{r} b_2 = U_e (S_f S_e)^{r} f $. The third entry in this column
is not needed, since $U_e b_3 = U_e U_{e'} f = 0$. Moreover, recall 
that $U_e b_1 = U_e e = e$, $U_e b_2 = U_e f = p e$ and $U_e b_4 = U_e U_{f'} e= (1 - p)^{2}e$. 
\\[0,5 cm]
$ U_e (S_f S_e)^{r} f  = \left(  - r + \frac{1}{4\sqrt{p(1-p)}} Im (\alpha_1^{r})   \right) U_e(b_1) $
\newline
\hspace*{3,5 cm}
$ + \frac{1}{2}\left(2 r + 1 + Re(\alpha_1^{r}) + \frac{1-2p}{2\sqrt{p(1-p)}} Im(\alpha_1^{r})\right)  U_e(b_2) $
\newline
\hspace*{3,5 cm}
$ + \frac{1}{2(1-p)} \left(2 r + 1 - Re(\alpha_1^{r}) - \frac{1-2p}{2\sqrt{p(1-p)}} Im(\alpha_1^{r})\right)  U_e(b_4)$
\newline
$= \left( - r + \frac{1}{4\sqrt{p(1-p)}} Im (\alpha_1^{r}) \right) e 
 + \frac{1}{2} \left( 2 r + 1 + Re(\alpha_1^{r}) + \frac{1-2p}{2\sqrt{p(1-p)}} Im(\alpha_1^{r}) \right)  p e$
\newline
\hspace*{3,5 cm}
$ + \frac{1}{2(1-p)} \left( 2 r + 1 - Re(\alpha_1^{r}) - \frac{1-2p}{2\sqrt{p(1-p)}} Im(\alpha_1^{r}) \right)  (1 -p)^{2}e$
\newline
$= \left( \frac{1}{2} - \frac{1-2p}{2} Re(\alpha_1^{r}) + \sqrt{p(1-p)}  Im(\alpha_1^{r}) \right) e$
\\[0,5 cm]
Therefore
$$ \mathbb{P}((S_f S_e)^{r}f|e) = \frac{1}{2} - \frac{1-2p}{2} Re(\alpha_1^{r}) + \sqrt{p(1-p)} Im(\alpha_1^{r}) $$
Since $\left| \alpha_1 \right| = 1$, $\alpha_1 = e^{it}$ 
with $t = arcsin (4(1-2p)\sqrt{p(1-p)})$. 
Furthermore, define $s := arcsin(2\sqrt{p(1-p)}$. 
Then $cos(s) = 1-2p $, 
since $(1-2p)^{2} + (2\sqrt{p(1-p)})^{2} = 1$, and 
\begin{align*}
\mathbb{P}(f|(S_e S_f)^{r}e) &= \frac{1}{2} - \frac{1}{2} \Bigl( cos(s) cos(rt) - sin(s) sin (rt) \Bigr) \\
&= \frac{1}{2} - \frac{1}{2} cos(s + rt)  \\
&= sin^{2} \Bigl( \frac{s + rt}{2} \Bigr) \\
&= sin^{2} \Bigl( (2r+1)arcsin(\sqrt{p}) \Bigr).
\end{align*}
The second and the third equality follow from the trigonometric identities 
$cos(x) + cos(y) = cos(x) cos(y) - sin(x) sin(y)$ 
and $sin^{2}(\frac{x}{2}) = \frac{1 - cos(x)}{2}$.
The last equality follows from the definitions of $s$ and $t$ and the following identity:
$$arcsin\left(2\sqrt{x - x^{2}}\right) + r \ arcsin\left(4(1-2x)\sqrt{x-x^{2}}\right) - (4r+2) arcsin\left(\sqrt{x}\right) = 0$$ 
by inserting $x = p$, which then gives $s+rt = (4r+2) arcsin(\sqrt{p}) $. 
This identity can be proved by differentiation with respect to $x$: 
The derivative is constantly zero and the function thus constant; checking the function for $x=0$
yields that it is constantly zero. \hfill $\Box$

\end{document}